\documentclass[aps,prl, showpacs,twocolumn,twoside,floatfix,a4]{revtex4}
\usepackage{hyperref}
\usepackage{bbm}
\usepackage{amsmath, amssymb}
\usepackage{color}
\usepackage{graphicx,epsfig}
\usepackage{pifont} 

\newcommand{\tick}{\ding{51}} % or 52
\newcommand{\cross}{\ding{53}} % or 54

\renewcommand{\>}{\rangle}
\newcommand{\Letter}{Letter\,}
\begin{document}

\title{Computational power of correlations}

\author{Janet Anders}
\email{janet@qipc.org}
\affiliation{Department of Physics and Astronomy, University College London, Gower Street, London WC1E 6BT, United Kingdom.}

\author{Dan E. Browne}
\email{d.browne@ucl.ac.uk}
\affiliation{Department of Physics and Astronomy, University College London, Gower Street, London WC1E 6BT, United Kingdom.}

\date{\today}

\begin{abstract}

We study the intrinsic computational power of correlations exploited in measurement-based quantum computation. By defining a general framework the meaning of the computational power of correlations is made precise. This leads to a notion of resource states for measurement-based \textit{classical} computation. Surprisingly, the Greenberger-Horne-Zeilinger and Clauser-Horne-Shimony-Holt problems emerge as optimal examples. Our work exposes an intriguing relationship between the violation of local realistic models and the computational power of entangled resource states.

\end{abstract}



\maketitle

\begin{figure}[t]
   \begin{center}
   \includegraphics[width=0.29\textwidth]{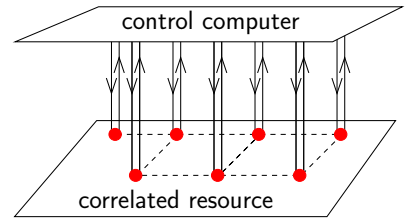}
       \caption{\label{fig:one-way} The control computer provides one of $k$ choices as the classical input (downward arrows) to each of the correlated parties (circles in the resource) and receives one of $l$ choices as the output. }
   \end{center}
\end{figure}

A striking implication of measurement-based quantum computation (MBQC) is that correlations possess intrinsic computational power. MBQC is an approach to computation radically different to conventional circuit models. In a circuit model, information is manipulated by a network of logical gates. In contrast, in the standard model of MBQC (also known as ``one-way'' quantum computation) information is processed by a sequence of adaptive single-qubit measurements on an entangled multi-qubit resource state \cite{MBQCMain,MBQCtutorials,Jozsa}. Impressive characterization of the necessary properties of quantum resource states that enable universal quantum computation in the measurement model has already been achieved \cite{CTNResource,GraphResource}. However, it is not the quantum states themselves, but the correlated classical data returned by the measurements which embodies this computational power. A necessary ingredient to extract this power is a classical {\em control computer} (see Fig.~\ref{fig:one-way}), which processes and feeds forward measurement outcomes and directs future adaptive measurements.  From this classical computer's perspective, the correlated measurement outcomes  enable it to compute problems beyond its own power. 

%%%%%%%%%%%%%%%%%%
%%%%%%%%%%%%%%%%%%

In this \Letter we will make the notion of the computational power of a correlated resource precise.  By doing so, a natural classical analogue of measurement-based computation emerges and we find a link to quantum non-locality.  Specifically, we show that the Greenberger-Horne-Zeilinger (GHZ) problem \cite{GHZ} and the Clauser-Horne-Shimony-Holt (CHSH) construction \cite{CHSH} emerge as closely related to  measurement-based classical computation (MBCC), as does the Popescu-Rohrlich  non-local box \cite{NLB}.

%\smallskip

{\it Framework for MBQC.--} We wish to study the computational power of correlated resources in a more general setting than the particular models of  MBQC which have been proposed \cite{MBQCMain,MBQCtutorials,Jozsa,CTNResource,GraphResource}. To achieve this, let us first define a general framework of computational models which shares the essential features of MBQC. It consists of two components, a correlated multi-partite resource and a classical control computer. A correlated multi-partite resource consists of a number of parties, which exchange classical information with the control computer, see Fig.~\ref{fig:one-way}. 
The correlations in their outputs are solely due to their joint history and no direct communication between parties is allowed during the computation. There shall be just a {\em single} exchange of data with each party. This restriction is an important assumption and we discuss its necessity and consequences in footnote \cite{endnote}. The party will receive an input from an alphabet of $k$ choices and will return one of $l$ outcomes.

The second component is a classical control computer of specified power. The control computer can store classical information, exchange it with the parties and compute certain functions. Notably, the classical control computer is the only part of the model where \emph{active} computation takes place. Before the computation commences, the system components are pre-programmed to specify the computation to be performed. Specifically, the control computer receives the functions it will evaluate and the individual parties receive a specific set of measurement bases,  or more generally a choice of $k$ settings. 

%\smallskip

This framework consists only of explictly \emph{classical} objects  - all quantum features are hidden in the possibly non-classical nature of the correlations. The framework captures the most general model of a single classical system (the control computer) interacting with multiple correlated (but  non-signalling) parties, with the key restriction that each party is addressed only once. However, we place as little restriction as possible on their internal structure. For example, the parties making up the system could be qubits, or physical objects of any dimension.  In fact, the framework is  so general that it admits models where the correlations between the parties do not obey quantum mechanics.

%Within the present framework, we now ask what sets of correlations are compatbile with quantum mechanics \cite{Tsirelson,IPSEntanglement}.

It is straightforward to see how the original (one-way) model fits into this framework. Each party holds a single qubit of the cluster state and a measuring device, pre-programmed with two sets of measurement bases 
$|0\rangle\pm e^{i\alpha}|1\rangle$ and 
$|0\rangle\pm e^{-i\alpha}|1\rangle$, where $\alpha$ is party-dependant and specific for a particular  computation. In this model $k=l=2$, i.e. only a single bit is sent to each party to specify the sign of the angle and the returned bit is the actual outcome of the measurement. It is remarkable that full universal quantum computation can be achieved with the minimal values of $k$ and $l$. Since this requirement represents the most challenging setting for a correlation to exhibit computational power, we adopt it for the remainder of this \Letter and leave non-binary communication for the discussion.

The starting point for our analysis is the observation that the control computer for a computation using the cluster state does \textit{not} require the full power of a universal classical computer. The only operations needed to control the measurements are parity calculations \cite{MBQCMain,Jozsa} which can be obtained with the logical XOR gate or, for a reversible scheme, with the CNOT gate. The \emph{parity computer} is  a device implementing circuits containing only CNOT operations and  NOT operations. It can solve a number of problems efficiently, such as calculating the parity of bit-strings, and simulating deterministic Clifford group quantum circuits (Gottesman-Knill theorem) \cite{Aaronson}. However, the parity computer is not able to calculate any unbalanced logical function, such as NAND, AND, OR or TOFFOLI. 

To denote the different degrees of computational complexity \cite{zoo} we will use the convenient notation established in computer science.  We shall only consider complexity classes which assume a polynomial computation time -- a physically realistic requirement. The computational power of the parity computer has been shown to lie in a complexity class named \emph{Parity-$L$}, or $\oplus L$ \cite{Aaronson,ParityL}, while universal classical and quantum computation are associated with classes $P$  and $BQP$ respectively. It is believed that $\oplus L$ is weaker than $P$  which, in turn, is weaker than $BQP$, however none of these inclusions are proven to be strict.

%%%%%%%%%%%%%%%%%%
\begin{table}[t]
	\begin{tabular}{|c|c|c|c|}
	\hline
	\hline
						& $\oplus L \to BQP $ 	& $P \to BQP$ 	& $\oplus L \to P$ \\ \hline	
		cluster states 		& \tick 				& \tick		& \tick		    \\
		lattice graph states \cite{GraphResource}		& \tick 				& \tick		& \tick		    \\
		certain CTN states \cite{CTNResource}		 & \cross$^?$			& \tick		& \cross$^?$  \\
		GHZ states\footnote{also (non-physical) Popescu-Rohrlich boxes} 
		& \cross 		& \cross		& \tick      \\  \hline \hline
	\end{tabular}
	\caption{\label{tab:resourcestates} The table indicates the computational power of the cluster state and other resource states. Cluster and graphs states are resource states promoting the parity computer to quantum universality ($\oplus L \to BQP$,  implying also $P \to BQP$ and $\oplus L \to P$). CTN states promote a universal classical control computer to a universal quantum computer ($P \to BQP$) while a polynomial supply of three-qubit GHZ states enables the parity computer to achieve full classical computation ($\oplus L \to P $). A cross (\cross) \ indicates that the resource is not capable of providing the specified computational enhancement, under the assumption that the complexity classes are distinct - i.e. $\oplus L \neq P \neq BQP$. A \cross$^?$ indicates a conjecture of this.}
\end{table}
%%%%%%%%%%%%%%%%%%

The notation $\oplus L \to BQP$ indicates that the parity computer is promoted to full quantum universality when, for example, the cluster state is used as the resource state. Other families of resource states are readily   classified within our framework, see Table~\ref{tab:resourcestates}. Two distinct groupings can be found in the literature. Graph states \cite{GraphResource,BKMP}, which employ solely the algebra of Pauli operators to ensure determinism, are in the class $\oplus L \rightarrow BQP$. Another family, the computational tensor network (CTN) states \cite{CTNResource}, enables universal measurement-based quantum computation via a different method of accounting for the random measurement outcomes. For some CTN states it is not possible to achieve the correction using Pauli operators only and addition modulo $n>2$ is employed.  ``Carrying'' in addition is equivalent to the AND operation and such arithmetic is not possible on the parity computer. Certain CTN states thus likely belong to a different class of computational power than the cluster states; specifically, being in class $P \rightarrow BQP$ but not $\oplus L \rightarrow BQP$ (indicated by \cross$^?$ in Table~\ref{tab:resourcestates}). This would also imply that $\oplus L \rightarrow P$ is not enabled by these CTN states. 

\smallskip

{\it Measurement-based classical computation.--} We now consider the reverse question: given the parity computer, what resource states can it be fed to raise its computational power? Adding any deterministic two-bit gate, which is not itself a product of NOT and CNOT operations, already constitutes a classical universal gate set. A resource state which promotes the parity computer to classical universality is a member of the class $\oplus L \to P$ \cite{BPPfootnote}, i.e. it enables measurement-based \emph{classical} computation (MBCC). It is clear that the cluster states (and any state in $\oplus L \to BQP$) belong to this class. However, it has been unresolved which features of the cluster state enable this computational enhancement and whether there exist states that enable $\oplus L \to P$ but not $\oplus L \to  BQP$. A way to promote the parity computer ($\oplus L$) to classical universality ($P$) is by giving it access to a polynomial number of universal gates, such as the NAND gate, see Tab.~\ref{tab:NAND}. One way to achieve this would be to take cluster states of a bounded size, each just large enough to  implement a NAND or TOFFOLI \cite{waterloo} via standard measurement patterns. Naturally, we wish to know how far the size of the resource can be reduced.

%%%%%%%%%%%%%%%%%%
\begin{table}[t]
	 \includegraphics[width=0.15\textwidth]{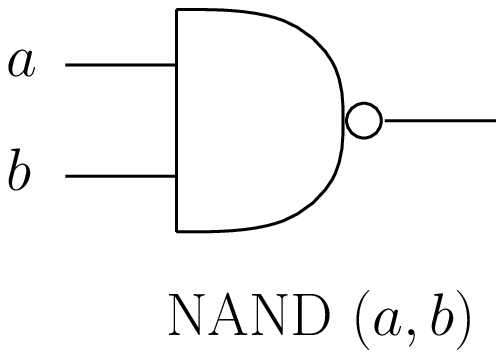} \hspace{0.22\textwidth} \phantom{0} \\
	\vspace{-0.13\textwidth}
	 \hfill
	\begin{tabular}{|cc|c|}
	\hline \hline
		$a$ & $b$ & NAND $(a, b)$\\ \hline	
		0 & 0 & 1\\
		0 & 1 & 1 \\
		1 & 0 & 1 \\
		1 & 1 & 0 \\ \hline \hline
	\end{tabular} \quad \quad \quad
	\caption{\label{tab:NAND} Pictogramm and truth table for NAND gate.}
\end{table}
%%%%%%%%%%%%%%%%%%

\smallskip

{\bf Theorem 1:} {\it There exists no \emph{bi-partite} quantum state upon which the parity computer can act to deterministically produce the NAND of two independent input bits.}

\smallskip

\begin{figure}[t]
	\begin{center}
	\includegraphics[width=0.17\textwidth]{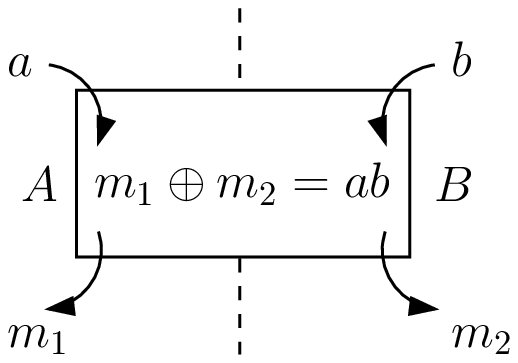} \quad \quad
	\includegraphics[width=0.21\textwidth]{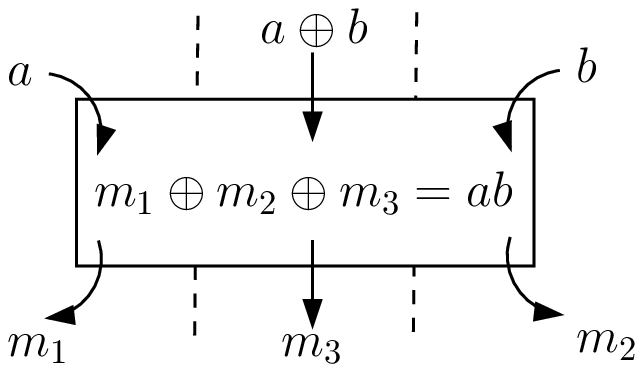}  
	\caption{\label{fig:PR-box} In this framework the non-local box and the three-qubit GHZ state have a strikingly similar structure. Traditionally, a non-local box (depicted on the left) is defined to implement the AND of the inputs $a$ and $b$. Measurements on the GHZ state, shown on the right, also implement an AND. In both cases the AND emerges as the parity of all outcomes, i.e. $\bigoplus_{j=1}^{n} m_j = a b =$ AND$(a,b)$; the difference is the number of parties, $n$, used. (The final negation of AND to NAND can be achieved by a single NOT operation by the parity computer.) }
	\end{center}
\end{figure}

{\bf Proof:} We prove this by contradiction. Assume that such a quantum state would exist. To satisfy the non-signalling condition and for the parity computer to be able to decode the result, the value of NAND of the input bits  $a$ and $b$ must be encoded in the parity of the two outputs $m_1$ and $m_2$, see Fig.~\ref{fig:PR-box}. Let $P_{(a,b)}$ be the probability of success of such a device acting on input $a$ and $b$. Since the gate is deterministic for all input values,
\begin{equation}
	\frac{1}{4}\sum_{{\{a,b\}\in\{0,1\}}}P_{(a,b)}=1
\end{equation}
must hold. This expression is a form of the CHSH quantity \cite{CHSH,vanDam}. Bell's theorem places the classical upper bound for this quantity at 0.75 and Tsirelson's bound \cite{Tsirelson} limits this quantity to $(2+\sqrt{2})/4\approx0.85$ for any correlations of  bi-partite quantum states. Thus, a bi-partite resource state for deterministically computing a NAND gate in this framework would  require correlations stronger than quantum mechanics. Indeed, the impossible device which implements this perfectly has been well-studied in the context of generalised no-signalling theories and is known as a Popescu-Rohrlich non-local box \cite{NLB}, see Fig.~\ref{fig:PR-box}.  Note that Tsirelson's bound (and thus this theorem) is valid for parties of  arbitrary dimension and internal structure. $\square$

\smallskip

{\bf Theorem 2:} {\it Measurements on a single \emph{three-qubit} GHZ state,  controlled by the parity computer,  enable the deterministic computation of the NAND gate.}

\smallskip

{\bf Proof:} The constructive proof follows directly from the well-known GHZ problem in the form introduced by Mermin \cite{GHZ}. Three measuring devices receive, respectively, the input bits $a,b, c \in \{ 0, 1\}$ and then act on three qubits which form a GHZ state, $|\psi\> = {|001\> - |110\> \over \sqrt{2}}$. The first two bits are independent, the third input $c=a\oplus b$ is fixed as the parity of the first two. Importantly, this operation can be performed by the controlling parity computer. Measuring devices which receive bit 0 measure Pauli observable $\sigma_x$, and those receiving 1 measure $\sigma_y$. The state  $|\psi\> $ is the only simultaneous eigenstate of the four equations corresponding to all four independent choices of input:
\begin{equation}\label{eq:stabilizer}
	\begin{split}
		\sigma_x \otimes \sigma_x \otimes \sigma_x |\psi \> = & -|\psi \> \\
		\sigma_x \otimes \sigma_y \otimes \sigma_y |\psi \> = & - |\psi \>\\
		\sigma_y \otimes \sigma_x \otimes \sigma_y  |\psi \> = & - |\psi \>\\
		\sigma_y \otimes \sigma_y \otimes \sigma_x  |\psi \> = & +|\psi \> .
	\end{split}
\end{equation}
Note, that in every case the eigenvalue is $(-1)^{\textrm{NAND} (a, b)}$. If we associate binary 0 with measured eigenvalue $+1$ and binary 1 with $-1$ and label the measurement outcome bits $m_{1}$,  $m_{2}$ and  $m_{3}$, respectively, Eqs.~\eqref{eq:stabilizer} guarantee that $m_{1}\oplus m_{2}\oplus m_{3}=$ NAND ($a, b$) \cite{Svetlichnyfootnote}. The parity computer can easily extract NAND ($a, b$) from the measurement outcomes $m_{j}  (j = 1, 2, 3)$ via a sequence of CNOT operations. $\square$

\smallskip

{\bf Corollary:} {\it A polynomial supply of three-qubit GHZ states is a resource for MBCC with deterministic gates which promotes the parity computer to classical universality ($\oplus L \rightarrow P$).  The GHZ states are optimal resources in that they \emph{minimize} the number of non-separable parties.}  

\smallskip

{\bf Proof:} Follows from the universality of NAND  and Theorems 1 and 2. $\square$

%%%%%%%%%%%%%%%%%%%
%%%%%%%%%%%%%%%%%%%

\smallskip

{\it Discussion.--} We have introduced a framework for classifying the computational power of correlations, and hence resource states, in measurement-based computation. The class of quantum states which enable deterministic universal classical computation when the control computer is the parity computer ($\oplus L\rightarrow P$) is particularly interesting.  We have shown that a polynomial supply of three-qubit GHZ states is an optimal resource for measurement-based classical computation, limiting the number of parties sharing entanglement to three. Generating the correlations of the NAND using only two correlated parties implies precisely the correlations of the non-local box, which violates the CHSH inequality maximally. Our framework thus unites the two most important ``non-locality paradoxes'', giving them a pleasing interpretation as computational tasks (see also \cite{vanDam}). Moreover, this equivalence delivers the simple explanation for the apparent violation of Tsirelson's bound for measurements on a GHZ state \cite{Cabello} when a non-standard definition of locality is employed. 

The introduced framework places the role of the classical data flow to the fore, leaving the internal structure of the parties entirely unrestricted. Alternative approaches are possible, for example, one could place restrictions on the allowed operations of the parties, such as the internal dimension or types of measurement allowed, and leave the classical data flow unrestricted. This could provide additional structure in the classification of resource states. In particular, permitting a higher degree of communication  (ie. $k,l>2$ ) would motivate the classification of the complexity of non-binary logic circuits with restricted gate sets. Such a complexity class might characterise the computational power needed for the control of measurements on certain CTN states, which appear to require non-binary modulo arithmetic.  Care would be needed to retain the correlation-based characteristic  of MBQC since active computation could take place within the individual parties, e.g. a party could hold a NAND gate, or even a full quantum computer. 

Finally, a notable feature of our results is that the measurements implementing the NAND can be made in parallel. The logical depth of a NAND-gate circuit implemented in this way will thus share the same scaling as the circuit implementation, with an additional factor due to parity calculations either side of the measurements. An alternative to implementing the circuit via measurements on multiple GHZ states would be to represent the whole logical circuit in the measurement outcomes of a single multi-qubit entangled state. This could imply new methods of circuit  parallelization via quantum means \cite{Parallelization}.

Aaronson and Gottesman  \cite{Aaronson} proved that Pauli measurements on any stabilizer state \ (such as a GHZ state) can be simulated in \emph{Parity-$L$}. This seems to contradict our result, if one assumes $\oplus L \neq P$, since we have shown that Pauli measurements on GHZ state enable universal classical computation. The important difference is that here the Pauli measurements needed are {\em adaptive}. The measurement made, $\sigma_x$ or  $\sigma_y$, is controlled upon the bit received by the measurement device. This is equivalent to a controlled-$\sqrt{Z}$ within the measurement device, which is not in the set of ``Clifford group'' operations considered in \cite{Aaronson}. In other words,  only non-adaptive fixed basis Pauli measurements on stabilizer states have been shown to lie in \emph{Parity-$L$}. This resolves the apparent contradiction.

The computational resource character of entangled states is a surprising feature of the quantum world. We hope that this \Letter\, helps to refine our understanding of this property and provides tools for its further analysis. This work reveals a number of open questions and underlines the important connections between physics and computer science which  quantum information science has been so successful in illuminating.

{\it Acknowledgements.--} We acknowledge inspiring and fruitful discussions with R. Blume-Kohout, H.J. Briegel, A. Broadbent, J. Fitzsimons, L. Hardy, E. Kashefi, D. Leung, A. Miyake,  S. Perdrix, B. Sanders, R. Spekkens, B. Toner, R. Werner and H. Wiseman. This work was funded by the EPSRC's QIPIRC programme and QNET network and the EU (QICS network).

%%%%%%%%%%%%%%%%%%%%%%


\begin{thebibliography}{99}

\bibitem{MBQCMain} 
	R. Raussendorf and H.~J. Briegel, Phys. Rev. Lett. {\bf 86},  5188  (2001); 
	R. Raussendorf, D.~E. Browne, and H.~J. Briegel, Phys. Rev. A {\bf 68},  022312  (2003).

\bibitem{MBQCtutorials}
	M.~A. Nielsen, Rep. Math. Phys. {\bf 57},  147  (2006);  
	D.~E. Browne and H.~J. Briegel, arXiv:quant-ph/0603226v2.

\bibitem{Jozsa} 
	R. Jozsa,  arXiv:quant-ph/0508124.

\bibitem{CTNResource}  
	D. Gross and J. Eisert,  Phys. Rev. Lett. {\bf 98},  220503  (2007); 
	D. Gross, J. Eisert, N. Schuch, and D. Perez-Garcia, Phys. Rev. A {\bf 76} 052315 (2007).
	
\bibitem{GraphResource} 
	M.~Van den Nest, A. Miyake, W. D\"{u}r, and H.~J. Briegel,  Phys. Rev. Lett. {\bf 97},  150504  (2006); 
	M.~Van den Nest, W. D\"{u}r, A. Miyake, and H.~J. Briegel, New. J. Phys. {\bf 9},  204  (2007).

\bibitem{GHZ} 
	D.~M. Greenberger, M.~A. Horne and A. Zeilinger, in: M. Kafatos (ed.), Bell's Theorem, Quantum Theory, and Conceptions, Kluwer Academic, Dordrecht, pp. 69 (1989); 
	N.~D. Mermin, American Journal of Physics {\bf 58},  731  (1990).

\bibitem{CHSH} 
	J.~S. Bell, Physics \textbf{1}, 195 (1964); 
	J.~F. Clauser, M.~A. Horne, A. Shimony, R.~A. Holt, Phys. Rev. Lett. \textbf{23}, 880 (1969). 

\bibitem{NLB} 
	S. Popescu and D. Rohrlich, Found. Phys. 24, 379Ð385 (1994); 
	V. Scarani, AIP Conference Proceedings {\bf 844}, 309 (2006), or  arXiv:quant-ph/0603017.
	
\bibitem{endnote}
If there were no limitation on the exchange of data between parties and control, the framework would admit many models of computation which do not share MBQC's correlation-based character, including the quantum circuit model. The correlations would then no longer be the sole source of computational power. 

\bibitem{Aaronson}  
	S. Aaronson and D. Gottesman, Phys. Rev. A {\bf 70},  052328  (2004).

\bibitem{zoo}
	The Complexity Zoo, edited by S. Aaronson and G. Kuperberg, \url{ http://www.complexityzoo.com/}

\bibitem{ParityL}  
	C. Damm, Inform. Process. Lett. {\bf 36},  247  (1990).

\bibitem{BKMP} 
	D.~E. Browne, E. Kashefi, M. Mhalla, and S. Perdrix, New. J. Phys. {\bf 9},  250 (2007).

\bibitem{BPPfootnote} Strictly speaking, the measurement outcomes also act as a resource of randomness, so the complexity class $BPP$ is  achieved.

\bibitem{waterloo} 	
	R. Blume-Kohout, A. Miyake, D. Leung, private communication.

\bibitem{vanDam} 
	W. van Dam, arXiv: quant-ph/0501159.

\bibitem{Tsirelson}
	B.~S. Tsirelson, %"Quantum Generalizations of Bell's Inequality",
	Lett. Math. Phys. {\bf 4}, 93 (1980).

\bibitem{Svetlichnyfootnote}
We note that no genuine three-party non-locality is required for these correlations because the corresponding Svetlichny inequality \cite{Slev} is not violated. In fact, a single non-local box can reproduce the correlations of the GHZ state with perfect probability \cite{Broadbent}. 

\bibitem{Slev}	
	G. Svetlichny, Phys. Rev. D {\bf 35}, 3066 (1987).

\bibitem{Broadbent}
	A. Broadbent, A. A. M\'{e}thot, Theor. Comp. Sci.,  \textbf{358} 3 (2006).

\bibitem{Cabello} 
	A. Cabello, Phys. Rev. Lett. \textbf{88} 060403  (2002).

\bibitem{Parallelization}
	A. Broadbent and E. Kashefi, arXiv:0704.1736 [quant-ph] (to be published in Theo. Comp. Sci.).

\end{thebibliography}
\end{document}